\documentclass{aastex62}
\shortauthors{N.T. Shukure, S.B Tessema, N. Gopalswamy.}

\begin{document}

\title{The Total Solar Irradiance variability in the Evolutionary Timescale and its Impact on the Mean Earth's Surface Temperature. }

\correspondingauthor{N.T. Shukure }
\email{nagessa@gmail.com, tessemabelay@gmail.com, nat.gopalswamy@nasa.gov}

\author[0000-0002-0786-7307]{N.T. Shukure}
\affil{Ethiopian Space Science and Technology Institute (ESSTI), 
Entoto Observatory and Research Center (EORC), \\
Astronomy and Astrophysics Research and Development Department, P.O.Box 33679 Addis Ababa, Ethiopia.}
\affil{Addis Ababa University
P.O.Box 1176 Addis Ababa, Ethiopia}
\affil{College of Natural \& computational Science, Dept. of Physics, Dilla University,
Box 419 Dilla, Ethiopia}

\author[0000-0002-0786-7307]{S.B Tessema}
 \affil{Ethiopian Space Science and Technology Institute (ESSTI), 
 Entoto Observatory and Research Center (EORC), \\
 Astronomy and Astrophysics Research and Development Department, P.O.Box 33679 Addis Ababa, Ethiopia.}
\affil{Addis Ababa University
P.O.Box 1176 Addis Ababa, Ethiopia}
\author[0000-0002-0786-7307]{N. Gopalswamy}
\affil{Solar Physics Laboratory, NASA Goddard Space Flight Center, Greenbelt, MD 20771, USA.}
\begin{abstract}
The Sun is the primary source of energy for the Earth.  The small changes in total solar irradiance (TSI)  can affect our climate in the longer timescale. 
In the evolutionary timescale, the TSI varies by a large amount and hence its influence on the  Earth's mean surface temperature  (T$_{s}$) also increases 
significantly.  We develop a mass-loss dependent analytical model of TSI  in the evolutionary timescale and evaluated its influence on the T$_{s}$. We 
determined the numerical solution of TSI  for the next 8.23 Gyrs to be used as an input to evaluate the   T$_{s}$ which formulated based on a zero-dimensional
energy balance model.  We used the present-day albedo and bulk atmospheric emissivity of the Earth and Mars as initial and final boundary conditions, 
respectively.  We found that the TSI increases by 10\% in 1.42 Gyr,  by 40\% in about 3.4 Gyrs, and by 120\% in about 5.229 Gyrs from now, while the T$_{s}$
shows an insignificant change in 1.644 Gyrs  and increases to 298.86 K in about 3.4 Gyrs. The T$_{s}$ attains the peak value of 2319.2 K as the Sun 
evolves to the red giant and emits the enormous TSI of 7.93$\times10^{6} Wm^{-2}$ in 7.676 Gys. At this temperature  Earth likely evolves to be a  liquid
planet. In our finding, the absorbed and emitted flux equally increases and approaches the surface flux in the main sequence, and they are nearly equal beyond
the main sequence,  while the flux absorbed by the cloud shows opposite trend.  

\end{abstract}



\section{Introduction} \label{sec:intro}

The Sun is the largest energy source to the Earth's atmosphere \citep{kren2015investigating,kren2017does}. The small changes in total solar irradiance (TSI),
which is also called solar constant ($S_{0}$), is the energy radiated by the Sun that is incident on the Earth's atmosphere.  TSI varies  by about 0.1\% over
the solar-cycle timescale
\citep{frohlich2006solar,kopp2016magnitudes}. It is varying with time and its tiny changes in TSI affect the Earth's climate in the longer timescale
\citep{eddy1976maunder}. This idea is  strengthened by many other studies 
\citep{haigh2007sun,leanjl2008hownaturalandanthropogenic,gray2010solar,ineson2011solar,ermolli2013recent,solanki2013solar,kopp2016magnitudes}. The 
space-borne instruments  have  been measuring the TSI since 1978. Its average value is 1361 $Wm^{-2}$ at 1 au \citep{kopp2011new}.\\  
\hspace*{0.5cm} 
In the evolutionary timescale, the Sun's luminosity changes by a large amount.The  term "evolution" refers a star's change over the course of time. Currently, the solar luminosity is increasing by 0.009\% per million years
\citep{hecht1994fiery}. At this rate,  the TSI increases by 0.1\% in 10 million years. Following this rate, TSI is expected to increase by 10\% in 1 Gyr 
at which time the Earth will become uninhabitable. In the same way, the luminosity will increase by 40\% in about 3.5 Gyrs
\citep{hecht1994fiery,kopp2016magnitudes}. As
\cite{kopp2016magnitudes} indicated, the TSI increases as the solar luminosity increases in the evolutionary timescale. However, there is a limitation of 
estimation of the TSI in the evolutionary timescale.  \cite{kasting1988runaway} estimated the 
TSI   to be varying from 1.1$S_{0}$ to 1.4$S_{0}$. This variation is in about 1.1 to 3.5 Gyrs as indicated by \cite{sackmann1993our,hecht1994fiery}. However, 
to study the influence of the TSI on the  Earth's mean surface temperature ($T_{s}$) during Sun's lifetime,  modeling of the TSI in relation to the evolutionary 
timescale is needed. \\
\hspace*{0.5cm} 
\cite{o2013swansong} used time-dependent luminosity to study the TSI variability and found its influence on $T_s$ in the evolutionary timescale. However, 
the solar mass-loss has an influence on the solar luminosity on this timescale. Moreover, their work is limited to the next 3 Gyrs, although further 
study is needed to understand the variability of an extended time interval. \cite{schroder2001solar} presented a model of  $T_{s}$ in which they include 
the effect of solar mass variation, and predict the $T_{s}$ of different planets including Earth over the lifetime of the Sun. They conclude that we can 
survive although the maximum ratio of the future to the present temperature is about 6.6. However, in their modeling, they assume that albedo ($\alpha$) 
is constant, and neglect the effect of bulk emissivity ($\varepsilon$). But,  $\alpha$  is expected to be changing, perhaps decreases in the future. 
Therefore, in our model, we incorporate the climate parameters: non-constant  $\alpha$ and $\varepsilon$. Moreover, in our modeling, we include the solar
 mass-loss in
both TSI and $T_{s}$ models.\\
\hspace*{0.5cm} 
The aim of this work is to model the TSI variability and  $T_{s}$ in the evolutionary timescale and to study the link between the two parameters.\\
\hspace*{0.5cm}
In this study, we developed an analytical model of the TSI variability and $T_{s}$ in the evolutionary timescale. The base of our formulation is the models
developed by \cite{shukure2019mass} and N. T. Shukure et al. (2021,  submitted to APJ). They developed a mass-loss dependent solar luminosity in the evolutionary timescale. To calculate
the solar mass-loss, they first calculate the solar mass reduction every 1 million years for the next 8.23 Gyrs. Then, they calculate the mass-loss  at each
interval of 1 million years by subtracting every newly reduced solar mass from the present-day solar mass.  Following this method, 
we calculate the mass-loss to model TSI in the evolutionary timescale. Using the newly formulated TSI model and zero-dimensional energy balance model
(EBM),
we formulate $T_{s}$. We used the numerical method to solve the newly formulated models by setting some boundary conditions to  $\varepsilon$ and  $\alpha$ 
that will be explained in detail in the method part. 

\section{Method}\label{mehod}
\subsection{Mathematical Formulation}
\subsubsection{The Solar Mass-loss}
In modeling TSI in evolutionary timescale, considering the solar mass-loss effect is mandatory. The solar mass reduction and solar mass-loss are modeled by
\cite{shukure2019mass}: 

\begin{equation}\label{newmass}
 M(t) = M_{\odot}e^{-\delta t},
\end{equation}
where $M_{\odot}$ is the present-day solar mass, $\delta$ determines the rate of decay \citep{Langtangen2016, Strang2021Exponential} of the total mass 
of the star. It is defined as the ratio of a star's mass-loss rate ($\dot{M}$) to the star's mass which is measured in per year ($yr^{-1}$) 
N. T. Shukure et al. (2021,  submitted to APJ). Eq. \ref{newmass} estimates the variation of the original mass of the Sun with time and t is the time needed 
to change the original mass of the star.
For $M(t=0) = M_{\odot}$ which is original mass of the Sun. However, the solar mass-loss $(\Delta{M})$ is the difference between original mass
$\&$ the remained mass $(M(t))$.\\
\begin{equation}\label{massloss}
 \Delta{M} = M_{\odot}-M(t).
\end{equation}
Eq. \ref{massloss}  estimates the solar $\Delta{M}$ at any time t.
\subsubsection{Luminosity-mass Loss Relation}\label{lum-mass}
 The  $\Delta{M}$ dependent luminosity that is  modeled by N. T. Shukure et al. (2021,  submitted to APJ) is: 
 \begin{equation}\label{Lm}
 L(\Delta{M}) = \frac{L_{\odot}}{1+\frac{2}{5}\left[1+\frac{1}{\delta t_{\odot}}ln\left(1-\frac{\Delta{M}}{M_{\odot}}\right)\right]-\frac{2}{5}},
 \end{equation}
where, $L_{\odot}= 3.85\times 10^{26}\:W$ is the present-day solar luminosity,  and $t_{\odot}$ = 4.57 Gyr   is the  present-day solar age
\citep{Dwivedi2010makes, feulner2012faint}. Eq. \ref{Lm} used as an input to model TSI in the evolutionary timescale.\\
\subsection{Modeling Total solar irradiance in the evolutionary timescale}
The TSI for the present-day solar constant  is \citep{o2013swansong}:
\begin{equation}\label{flu}
 S_{0} = \frac{L_{\odot}}{4\pi K d^{2}_{E}},
\end{equation}
where k is a constant (k $\approx$ 1) and $d_{E}$ is the Earth-Sun distance. Eq. \ref{flu} estimates the amount of energy which cross an area $A = 1m^{2}$
(one square meter). 
For  our new model of luminosity in the evolutionary timescale, 
Eq. \ref{flu} can be re written as:
\begin{equation}\label{flu1}
 S_{evolution} = \frac{L(\Delta{M})}{4\pi K d^{2}_{E}}.
\end{equation}
Re writing Eq. \ref{flu1} in terms of Eq. \ref{Lm} become
\begin{equation}\label{flu2}
 S_{evolution} = \frac{S_{0}}{1+\frac{2}{5}\left[1+\frac{1}{\delta t_{\odot}}ln\left(1-\frac{\Delta{M}}{M_{\odot}}\right)\right]-\frac{2}{5}}.
\end{equation}
Eq. \ref{flu2} estimates the   TSI variability in the evolutionary timescale as solar mass-loss significantly increases. 
\subsection{ Earth's Mean Surface Temperature Variation}
 The temperature of any planet varies as the Sun evolves. Our main interest in this section is to model  $T_{s}$. In the modeling, we excluded the greenhouse 
 effect on the $T_{s}$ variation. The base of our modeling is the EBM. The energy flux that crosses a unit area is estimated by Eq. \ref{flu}. If we consider
 the region of our Earth that faces the Sun as a circular area, the total amount of energy flux ($\phi$)  that intercept the Earth's circular area  is the 
 product of the flux through the unit area ($S_{0}$) and the area of the circle: 
 \begin{equation}\label{intercept}
   \phi_{intercept} = S_{0} \pi R^{2}_{E},
 \end{equation}
 where $R_{E}$ is the radius of Earth. 
 However, the total surface area of Earth $(A_{E})$ is  
  \begin{equation}
 A_{E} = 4\pi R_{E}^{2}.
 \end{equation}
 The average energy absorbed  by the Earth is the ratio of the total energy absorbed by  the Earth to the surface area of the Earth and it is written as:
 \begin{equation}\label{absorbed1}
  \phi_{absorbed} = \frac{S_{0}}{4}.
 \end{equation}
 Eq. \ref{absorbed1} can be the flux absorbed by a completely dark
planet whose albedo is zero. However,  planet Earth has an albedo different from zero. Hence, 
 the absorbed $\phi$  by Earth is defined as \cite{Mcguffie1987ACM}:
 \begin{equation}\label{flux_absorbed}
  \phi_{absorbed} = \frac{S_{0}}{4}(1-\alpha), 
 \end{equation}
 where $\alpha$ is the temperature-dependent Earth's albedo taken to be varying in the interval $0 \le \alpha \le 1$.
  Energy  emitted by Earth's surface is estimated by the Stefan-Boltzmann law:
 \begin{equation}\label{surface_flux}
  \phi_{s} = \sigma T_s^{4},
 \end{equation}
where $\phi_{s}$ is the flux that escapes from the surface of the Earth, $\sigma$ is Stefan-Boltzmann constant and  $T_{s}$ is the Earth's mean surface 
temperature.
However, Eq. \ref{surface_flux} 
needs a correction for the cloud absorption.The flux absorbed by clouds is:
\begin{equation}\label{emitted}
  \phi_{c} = \varepsilon\sigma T_s^{4},
 \end{equation}
 where $\phi_{c}$ flux absorbed by cloud, $\varepsilon$ is dimensionless bulk emissivity that can  be in the interval $0 \leq \varepsilon < 1 $ 
 \citep{swift2018new}. 
 The flux radiated  out of the atmosphere is :
 \begin{equation}\label{corr_emitted}
  \phi_{emitted} = \left(1-\varepsilon\right)\sigma T_s^{4}.
 \end{equation}
 When the energy of a planet is at thermal equilibrium, the absorbed and emitted  $\phi$ by Earth become
 equals.
 \begin{equation}\label{equilibrum}
  \phi_{absorbed} = \phi_{emitted}
 \end{equation}
Substituting Eq. \ref{flux_absorbed} and \ref{corr_emitted} in to Eq. \ref{equilibrum} one  can calculate the $T_{s}$ as:
 \begin{equation}\label{TEo}
  T_s = \left(\frac{S_{0}(1-\alpha)}{4\left(1-\varepsilon\right)\sigma}\right)^{\frac{1}{4}}.
 \end{equation}
Eq. \ref{TEo} estimates the expected present-day  Earth's mean temperature. In the evolutionary timescale, the TSI is expected to be changing
(i.e., $S_{0}\rightarrow S_{evolution}$) and hence, 
it influences  $T_{s}$. Therefore, the expected $T_{s}$ in the evolutionary timescale can be estimated 
by rewriting Eq. \ref{TEo} in terms of  $S_{evolution}$.
\begin{equation}\label{Tevol}
  T_s = \left(\frac{S_{evolution}(1-\alpha)}{4\left(1-\varepsilon\right)\sigma}\right)^{\frac{1}{4}},
 \end{equation}
 where $T_s$  changes as the Sun evolves. Since the Earth's
 surface covered by ice will reduce in the future, which will result in the decrease of Earth's albedo. Substituting Eq. \ref{flu2} in to \ref{Tevol},
 $T_{evolution}$ is estimated as:
 \begin{equation}\label{Tfinal}
  T_{evolution} = \left(\frac{(1-\alpha)S_{0}}{4\left(1-\varepsilon\right)\sigma\left(1+\frac{2}{5}\left[1+\frac{1}{\delta t_{\odot}}ln\left(1-\frac{\Delta{M}}{M_{\odot}}\right)\right]-\frac{2}{5}\right)}\right)^{\frac{1}{4}}
 \end{equation}
 Eq. \ref{Tfinal} predicts the $T_s$ of our planet in the evolutionary timescale. 
\subsection{Numerical Computations of the Model }
In the numerical analysis, we assumed that the present-day position of the Earth's orbit is not changing as the Sun evolves. We numerically solved
Eq. \ref{flu2} by  setting $\delta$ to be varying as 
$10^{-13}yr^{-1}\leq\delta \leq 6\times10^{-11}yr^{-1}$ based on the present-day value of $\dot{M} =  10^{-13}M_{\odot}yr^{-1}$ reported by 
\cite{feulner2012faint}.
We also  assumed value of $\dot{M} = 6 \times 10^{-11}M_{\odot}yr^{-1}$ during red giant phase
and we used the solar observables to be $S_{0} = 1361 Wm^{-2}$ \citep{schmutz2013total}, $M_{\odot} = 1.99\times10^{30}$ \citep{kaplan1981iau} 
and the  present-day solar age as initial 
time $t_{\odot} = 4.57\times10^{9}$ \citep{bahcall1995solar, feulner2012faint}. The solar mass reduction 
is calculated for about 8.23 Gyrs from now. The new 
mass of the Sun, M(t) is recorded at the intervals of  every $1Myrs$  
based on Eq. \ref{newmass} to have a total of  8231 numerical values.  The total numerical values are used  for further
analysis.\\
\hspace*{0.5cm}
similarly,  the solar $\Delta{M}$  in the same time interval is calculated using Eq. 
\ref{massloss}.  Each of the new numerical values is subtracted from the present-day mass of the Sun in each time step ($1Myrs$). The new values of
$\Delta{M}$ are
used as an input to calculate the TSI  in the same time intervals.\\ 
\hspace*{0.5cm}
To solve Eq. \ref{Tfinal} we set some boundary conditions of  $\alpha$ and $\varepsilon$. We set the present-day values of the Earth's $\alpha$ and
$\varepsilon$ to be 0.3  and 0.407, respectively, as initial boundary conditions. By assuming the future Earth's atmosphere to be the present-day 
atmosphere of Mars, we set the final boundary condition to be 0.250 and  0.0038, respectively, for  $\alpha$ and $\varepsilon$ 
\citep{swift2018new}. \\
\hspace*{0.5cm}
We numerically solved  Eq. \ref{Tfinal} for the next 8.23 Gyrs by using the numerical solution of TSI as an input. Our program determines the $T_{s}$ 
every 1 million years.  We then calculate  $\phi_{absorbed}$, $\phi_{s}$, $\phi_{c}$, and $\phi_{emitted}$ based on Eq. \ref{flux_absorbed},
\ref{surface_flux}, \ref{emitted}
, and \ref{corr_emitted}, respectively,  by using the $T_{s}$ solution as an input.   
\section{Result and Discussion}
\subsection{Total solar irradiance variability in the evolutionary timescale}\label{TSIe}
\begin{figure}[h!] 
 \centering
 \includegraphics[width=18cm,height=6cm]{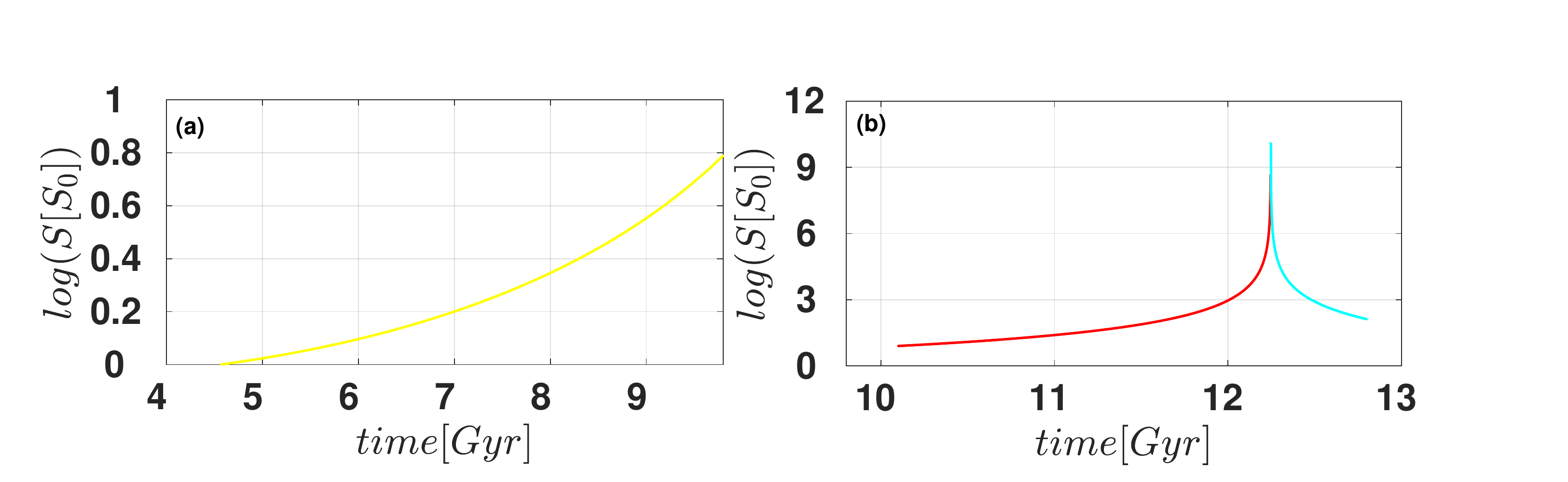}
 \caption{Evolution of TSI. (a) on the main sequence. (b) beyond the main sequence produced based on Eq.\ref{flu2}; the red line is to indicate before the Sun 
 loses its hydrogen envelope and cyan line after the Sun loses its hydrogen envelope.}
 \label{fig:model}
\end{figure}

Figure \ref{fig:model} shows the TSI variability as the Sun evolves.  Figure \ref{fig:model}a indicates that  the TSI increases to 1448.1 $Wm^{-2}$
in the first 1 Gyr from now, representing an increase by 6.4\% with respect to the present-day value.  At about 1.42 Gyr from now,
the TSI changes to 1497.1 $Wm^{-2}$ which is about 10\% increase compared to the present value. At this value of TSI, the Earth will lose water 
\citep{hecht1994fiery}.
However according to our model, the 10\% variability achieved at about 1.42 Gyrs that lags by 0.32 Gyrs as compared with Hecht's finding.\\
\hspace*{0.5cm}
In 3.374 Gyrs, the TSI increases to 
1905.4 $Wm^{-2}$, and it is about 40\% increment. This is the  \cite{kasting1988runaway} flux at which an ocean evaporates entirely. This value achieved at
about
3.5 Gyrs as indicated by  \cite{sackmann1993our,hecht1994fiery}. In our model, the 40\% increment of the TSI occurs before 3.5 Gyrs.\\
\hspace*{0.5cm}
At the end of the time in the main sequence,   TSI increases to 2994.2 $Wm^{-2}$ at about 5.23 Gyrs from now. As compared to $S_{0}$ it increased by about 120\%. 
Beyond the main sequence, the TSI increases enormously.  Figure \ref{fig:model}b shows the peak value of 7.93$\times10^{6}$ $Wm^{-2}$  at about 
an age of 7.676 Gyrs from now. This age is the age at which the Sun approaches the final stage of   RGB \citep{sackmann1993our}. 
N. T. Shukure et al. (2021,  submitted to APJ) indicated that the present-day Sun evolves into the white dwarf of mass 0.46 $M_\odot$ in the next 8.23 Gyrs. At 
this age, the dwarf
will have the TSI of -1.13$\times10^{4}$ $Wm^{-2}$. The negative sign indicates the TSI  is below the present-day value of the Sun. 
\subsection{ Earth's Mean Surface Temperature Variation in Evolutionary Timescale} 
\begin{figure}[h!]
 \centering
 \includegraphics[width=18cm,height=6cm]{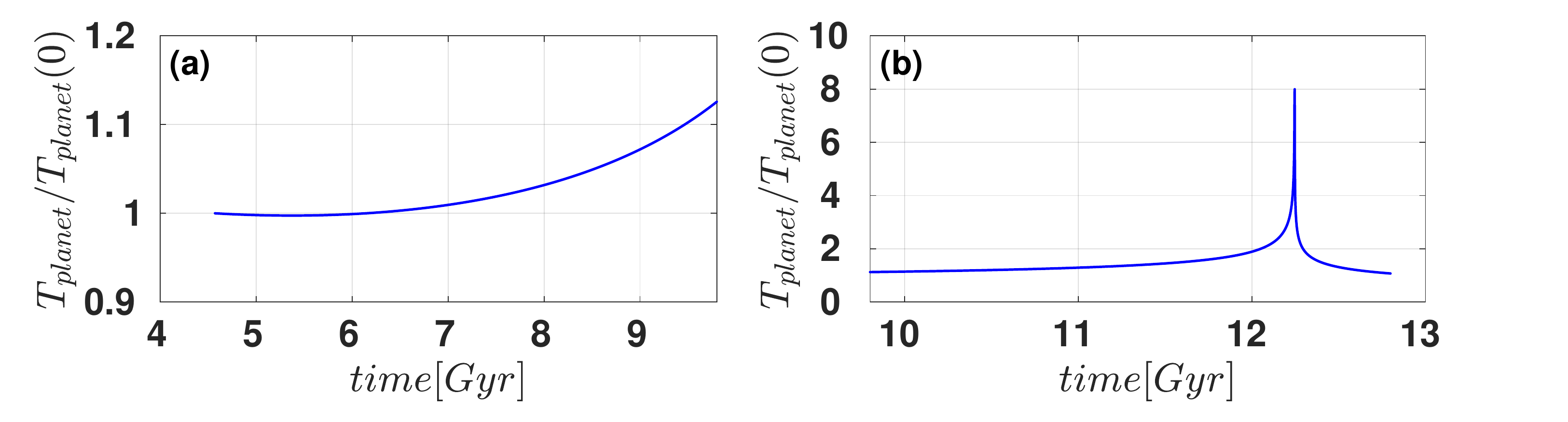}
 \caption{ $T_{s}$  relative to its present temperature which is produced based 
 on Eq. \ref{Tfinal}. (a) The predicted variation for the next 5.23 Gyrs of the Sun's life, up
to the end of the main sequence. (b) On the  expansion of the Sun   on the RGB . The maximum ratio is
about 7.99.}
 \label{fig:Tplanet}
\end{figure}

It is known that the present-day $T_{s}$   is  288 K \citep{schroder2001solar,Mcguffie1987ACM}. This temperature is expected to  change during the Sun's
evolution. Figure \ref{fig:Tplanet}a shows the $T_{s}$ variation as the Sun  evolves in the main sequence and Figure \ref{fig:Tplanet}b for beyond main 
sequence produced based on Eq.\ref{Tfinal}.  The equation predicts the present-day $T_{s}$ to be 290.16 K. This 
value is close to the experimental mean value.\\
\hspace*{0.5cm}
We assumed that $\alpha$ and $\varepsilon$ are decreasing up to the present-day values of Mars. Our basic assumption is that in the future, 
$\alpha$ and $\varepsilon$ of Earth are  decreasing. But it is difficult to estimate the final limit. We select the present-day value of Mars's 
$\alpha$ and $\varepsilon$. This is because the present-day atmosphere of Mars has no right conditions. The right condition for human life is
in the temperature range between  270 K and 300 K on the Earth. According to the model of \cite{schroder2001solar}, the range of temperature ratio and times 
for which human life might survive on Mars is 1.29 - 1.43 and 11.6 - 11.7 billion years, respectively. This indicates that  about 7 billion years are needed 
to warm Mars. However, human life  might be extinct from  the Earth before Mars is warm enough:  oceans would evaporate entirely in 3.5 Gyrs from now as 
estimated by \cite{sackmann1993our} and in 3.374 Gyrs by N. T. Shukure et al. (2021, submitted to APJ).  Hence, its $\alpha$ and 
$\varepsilon$ assumed to be  decreasing. 
Therefore,  we select its initial value as our final boundary condition by assuming that the parameters decrease in the selected time boundary. The Eq.\ref{Tfinal} predicts that the $T_{s}$ does not show a 
significant change in the absence of greenhouse effect for the next 1.644 Gyrs. This shows the influence of  TSI in the evolutionary timescale on $T_s$ is 
not significant although the TSI increased to 1528.1 $Wm^{-2}$. Including the effluence of human activities and the greenhouse effect my change the 
magnitude of temperature in this interval which results in the Earth losing its water \citep{kasting1988runaway}.\\
\hspace*{0.5cm}
Beyond 1.644 Gyrs from now,  the temperature of Earth is observed to be rising significantly (see Figure
\ref{fig:Tplanet}a). This is perhaps due to the increment of moisture in the Earth's atmosphere that induces an extra temperature which influences the $T_{s}$ 
\citep{o2013swansong}.  At about 3.374 Gyrs, when TSI increases to 1905.4 $Wm^{-2}$ (40\% increment from present-day value ),  part of this energy   absorbed 
by the  Earth's atmosphere raises the $T_{s}$ to 298.86 K. According \cite{kasting1988runaway},  the oceans would evaporate entirely when the TSI become 
 1905.4 $Wm^{-2}$.  $T_{s}$ rises by 3\% as predicted by Eq.\ref{Tfinal}. The equation also predicts the $T_{s}$  to be 326.72 K  in 5.23 Gyr from now when
 the Sun completes its main sequence life. This happens when the TSI becomes 2994.2 $Wm^{-2}$ (increases by 120\%).\\
 \hspace*{0.5cm}
 If we consider the assumption of \cite{schroder2001solar}, that all forms of life will extinct when $T_{s}$ has reached 380 K.  Eq.\ref{Tfinal} predicts
 that this will happen at the age of 7.13 Gyrs  (see Figure \ref{fig:Tplanet}b). This shows our equation predicts this phenomenon happens 0.1 Gyrs before 
 the Schr\"{o}der et al., prediction. This is because, in our model, we included $\varepsilon$ and $\alpha$  parameters and let them vary.  However, humans 
 will have to leave much sooner, similar to the present-day situation, perhaps before the model temperature reaches 291.9 K in 2.2 Gyrs, as our model predicts. 
 Figure \ref{fig:Tplanet}b also shows that $T_{s}$ attains a peak value of 2319.2 K, where the ratio is 7.7  as the Sun evolves to
 the red giant phase. This value is much greater than the melting points of rocks and minerals as investigated by  \cite{eppelbaum2014thermal}. This finding
 indicates the bad news that our planet Earth will have the probability of disappearance or have a chance of evolving into the liquid planet
 due to extreme  temperature increment in the next 7.676 Gyrs
 during Sun's red giant stage. However, beyond this age the good news is since the Sun's energy output observed to be decreasing, the $T_{s}$ also decreases
 and becomes 311.34 K at the end of 8.23 Gyrs from now. The decrease in the temperature of the planet may have allowed our planet to revert to a solid state. 

\subsection{Earth's flux variation surface to top of atmosphere}
\subsubsection{Variation in the main sequence}
So far we discussed the $T_{s}$ variation due to solar evolution. Since the magnitude of an energy flux entering  into
Earth's atmosphere is expected to vary with the altitude of our atmosphere, we estimate the energy flux at different levels by assuming that $T_{s}$ is 
independent of the atmospheric altitude.\\
\hspace*{0.5cm}
Figure \ref{fig:fluxes}a shows the energy flux variation at different levels of Earth's atmosphere. The $\phi_{absorbed}$ and $\phi_{emitted}$  that are
produced based on 
Eq. \ref{flux_absorbed} and \ref{corr_emitted}, respectively,  are  equal in the figure. Their present-day value is 238.41 $Wm^{-2}$ that shows the energy
balance of the system. The two
values increase as the Sun evolves in the main sequence, and approach the values of $\phi_{s}$  whose present-day value is 401.91 $Wm^{-2}$  based on
Eq. \ref{surface_flux}. This value is shown to be greater than  $\phi_{absorbed}$ since it is dependent only on the $T_{s}$ of the Earth. The $\phi_{s}$ 
does not show significant change for the next 2.83 Gyrs.  If the $T_{s}$ varies according to the prediction of our model, the life on the Earth's surface
would have a probability of surviving up to 2.83 Gyrs from now. This result is close to that of \citet{o2013swansong}, which predicts the maximum lifetime
for the life on our planet is 2.8 Gyrs from now. The $\phi_{c}$ is found to be decreasing from the present-day values of 163.58 $Wm^{-2}$. This is because 
we assumed the cloud emissivity to be decreasing until it becomes equal to the present-day value of  Mars. The values of the fluxes at different time intervals
are illustrated in Table \ref{Tab:sample}.\\

\begin{table}
\begin{center}
\caption{ The sample of the impute variables to calculate the TSI and mean Earth's Surface temperature variation and their sample result. }\label{Tab:sample}
\fontsize{8}{12}\selectfont
 \begin{tabular}{l l l l l l l l l l}
 
  \hline\noalign{\smallskip}
Age &  TSI & $T_{s}$&$\phi_{absorbed}$ & $\phi_{c}$ & $\phi_{s}$ &$\phi_{emitted}$ & $\alpha$&$\varepsilon$\\
$(Gyrs)$ & $(Wm^{-2})$ & $(K)$ & $(Wm^{-2})$ & $(Wm^{-2})$ & $(Wm^{-2})$&$(Wm^{-2})$&--&--\\
  \hline\noalign{\smallskip}
4.570  & 1361.908 &  290.160&  238.334 &  163.578&  401.912&238.334&0.300 &  0.407 \\ 
5.990&   1497.552&   289.889&  265.301& 135.113&  400.414&  265.301&  0.291& 0.337\\
7.944&  1905.465&  298.908&   343.221&  109.400&  452.621&  343.221&  0.280&   0.242      \\
9.800&  2995.261&   326.615&   547.963&   97.287&   645.251&   547.963&   0.268&   0.150\\
11.700& 12089.574& 452.832&   2246.597&   137.543&   2384.139&   2246.597&   0.257&   0.058 \\
12.246& 7925489.4& 2277.903& 1479360.5&   47234.825&   1526595.300&   1479360.500&   0.253&   0.031\\
12.247&  -34065380.000&   2319.196&  -6358646.800&  -202694.910&  -6561341.700&  -6358646.800&   0.253&   0.031\\
12.800&  -11308.673&   311.247&  -2120.376&-8.088&  -2128.464&  -2120.376&   0.250&  0.004\\
  \noalign{\smallskip}\hline
\end{tabular}
\end{center}
\end{table}
\subsubsection{Variation  beyond the main sequence}
\begin{figure}[h!]
 \centering
 \includegraphics[width=18cm,height=12cm]{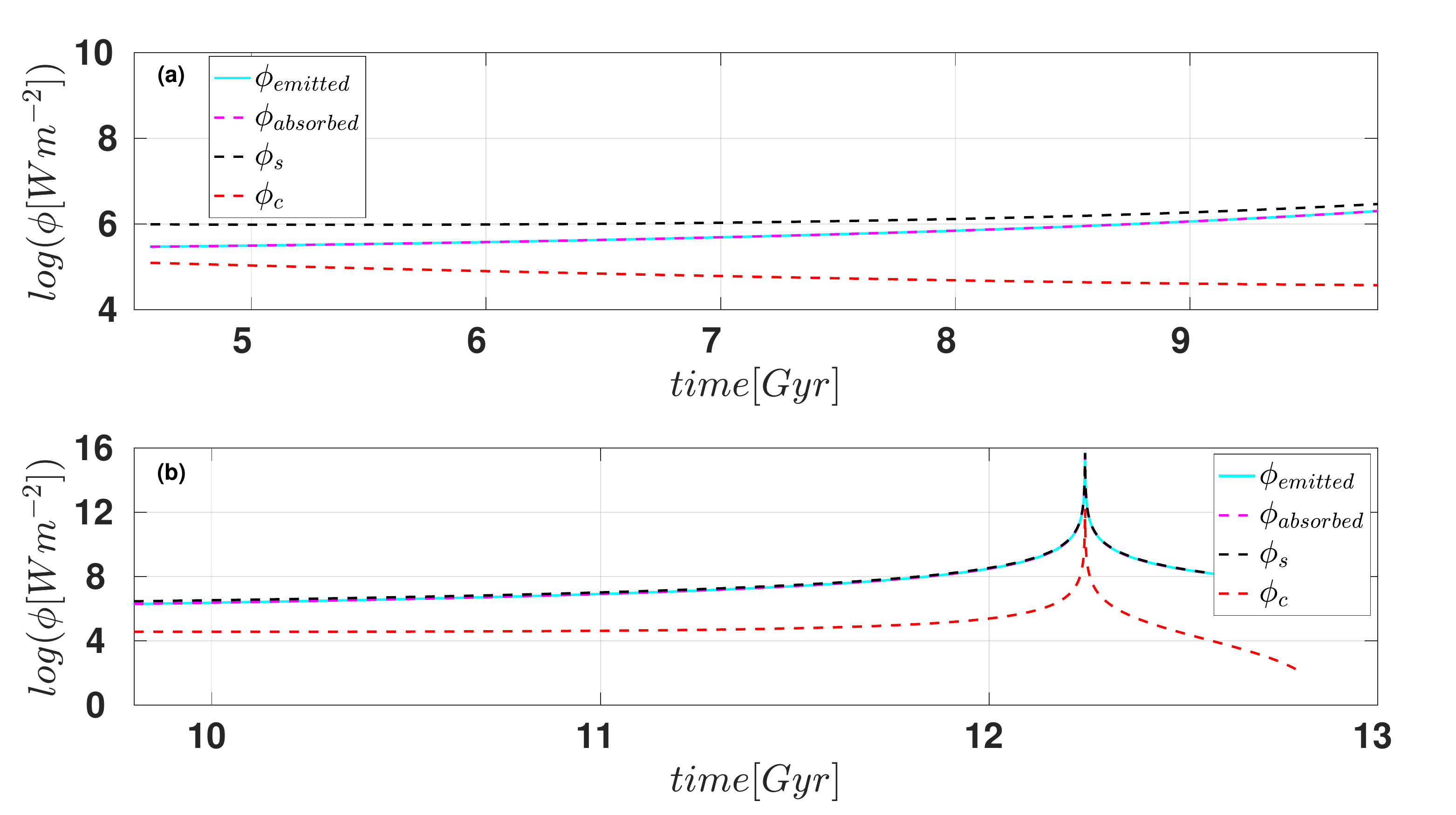}
 \caption{The Earth's flux variations from surface to top of the atmosphere. Panel (a) during the main sequence and (b) beyond  the main sequence. 
 }
 \label{fig:fluxes}
\end{figure}

The variation of various energy fluxes in the Earth's atmosphere beyond the main sequence is shown in  Figure \ref{fig:fluxes}b. Unlike in the main 
sequence, 
all fluxes except $\phi_{c}$  have almost the same values. For example, at about 5.43 Gyrs, $\phi_{absorbed}$, $\phi_{emitted}$, and $\phi_{s}$ has   the 
values of 590.97 $Wm^{-2}$, 590.97 $Wm^{-2}$, and  687.96 $Wm^{-2}$, respectively. However, $\phi_{c}$ have the value of 96.99 $Wm^{-2}$. At about 12.25 Gyrs,
when the Sun gets its peak expansion, all the fluxes attain their peak values as indicated in the figure. Additional values
of  
the fluxes at different time intervals are shown in the Table \ref{Tab:sample}.\\  

\section{Conclusion}
 We modeled the  TSI variability in the evolutionary timescale. It shows a significant influence on the  Earth's mean surface temperature. The following are 
 the 
 conclusions from the above analysis.
 \begin{enumerate}
  \item  The TSI varies by 6.4\% in 1 Gyr increasing from 1361 $Wm^{-2}$ to 1448.1 $Wm^{-2}$
  \item The TSI increases to 1497.1 $Wm^{-2}$ (by 10\%) from a present-day value in 1.42 Gyr from now.
  \item The  TSI increases by 40\% at the end of  3.4 Gyrs from now. 
  \item The model also predicts the TSI variability when the Sun is at the end of the main sequence ton be about 2994.2 $Wm^{-2}$ at about 5.23 Gyrs from now.
  \item The  Earth's mean surface temperature does not change significantly with a value of  290.16 K at  1.644 Gyrs from now.
  \item At about 3.4 Gyrs, when TSI increases to 1905.4 $Wm^{-2}$, the Earth's surface temperature becomes 298.86 K.
  \item The maximum Earth's temperature is predicted when the Sun becomes an RGB to be  2319.2 K turning Earth into a  liquid planet. 
  \item  The absorbed flux and emitted flux observed to be increasing equally and approach the surface flux at the end of the main sequence, 
  while the flux absorbed by clouds is found to be decreasing.
  \item Beyond the main sequence, the surface flux, absorbed flux and emitted flux become nearly equal attaining a  peak value at about 7.676 Gyrs
  from now. However, the flux absorbed by the clouds decreases relative to others, although it peaks at the same points as other fluxes.  
 \end{enumerate}
\section*{Acknowledgments} 
N.T.S thanks Entoto Observatory and Research Center (EORC) for hosting his research. 
This research was partially supported by the International Science Program (ISP) under the East African Astrophysics Research Network (grant
code IPPS/AFRO5).  N.T.S also thanks Dilla University for covering the living expenses in Ethiopia NG is supported by NASA's Living With a Star Program.



\end{document}